\RequirePackage[left]{lineno}
\documentclass[twocolumn,superscriptaddress,showpacs,prl]{revtex4}
\usepackage{graphicx}
\usepackage{dcolumn}
\usepackage{natbib}
\usepackage{color}
\usepackage{float}
\usepackage[abs]{overpic}
\newcommand{\kggg}{K^0_L \rightarrow 3\gamma}

\newcommand{\kpp}{K^0_L\rightarrow\pi^0\pi^0}
\newcommand{\kppp}{K^0_L\rightarrow 3\pi^0}
\newcommand{\kgg}{K^0_L \rightarrow \gamma \gamma}

\newcommand{\ses}{(3.23\pm0.14)\times10^{-8}}

\newcommand*{\TAIWAN}{%
Department of Physics, National Taiwan University, Taipei, Taiwan 10617 Republic of China}
\newcommand*{\PUSAN}{%
Department of Physics, Pusan National University, Busan, 609-735 Republic of Korea}
\newcommand*{\SAGA}{%
Department of Physics, Saga University, Saga, 840-8502 Japan}
\newcommand*{\DUBNA}{%
Laboratory of Nuclear Problems, Joint Institute for Nuclear Research, 
Dubna, Moscow Region, 141980 Russia}
\newcommand*{\SOKENDAI}{%
Department of Particle and Nuclear Research, 
The Graduate University for Advanced Science (SOKENDAI), Tsukuba, Ibaraki, 305-0801 Japan}
\newcommand*{\KEK}{%
Institute of Particle and Nuclear Studies, 
High Energy Accelerator Research Organization (KEK), Tsukuba, Ibaraki, 305-0801 Japan}
\newcommand*{\OSAKA}{%
Department of Physics, Osaka University, Toyonaka, Osaka, 560-0043 Japan }
\newcommand*{\YAMAGATA}{%
Department of Physics, Yamagata University, Yamagata, 990-8560 Japan}
\newcommand*{\CHICAGO}{%
Enrico Fermi Institute, University of Chicago, Chicago, Illinois 60637, USA }
\newcommand*{\NDA}{%
Department of Applied Physics, National Defense Academy, Yokosuka, Kanagawa, 239-8686 Japan}
\newcommand*{\RCNP}{%
Research Center of Nuclear Physics, Osaka University, Ibaraki, Osaka, 567-0047 Japan}
\newcommand*{\KYOTO}{%
Department of Physics, Kyoto University, Kyoto, 606-8502 Japan}
\newcommand*{\IHEP}{%
Institute for High Energy Physics, Protvino, Moscow region, 142281 Russia}
\newcommand*{\GOMEL}{%
Scarina Gomel' State University, Gomel', BY-246699, Belarus}
\newcommand*{\ARIZONA}{%
Department of Physics and Astronomy, Arizona State University, Tempe, Arizona, USA}
\newcommand*{\RIKEN}{%
RIKEN SPring-8 Center, Sayo, Hyogo, 679-5148 Japan}
\newcommand*{\NY}{%
University  of Rochester, Rochester, NY 14627}
\newcommand*{\CERN}{%
CERN, CH-1211 Gen\`{e}ve 23, Switzerland}
\begin{document}
\date{November 19 2010}
%================================================
%   Title 
%================================================
\title{
Search for the decay $\kggg$}

%======================================================================
%   Author lists : 
%      Duplicated from Jon's K->pi0pi0nunu paper
%======================================================================
\author{Y.~C.~Tung}\affiliation{\TAIWAN}
\author{Y.~B.~Hsiung}\affiliation{\TAIWAN}
\author{J.~K.~Ahn}\affiliation{\PUSAN} 
\author{Y.~Akune}\affiliation{\SAGA} 
\author{V.~Baranov}\affiliation{\DUBNA}
\author{K.~F.~Chen}\affiliation{\TAIWAN}
\author{J.~Comfort}\affiliation{\ARIZONA} 
\author{M.~Doroshenko}\altaffiliation{Present address: \DUBNA}\affiliation{\SOKENDAI} 
\author{Y.~Fujioka}\affiliation{\SAGA} 
\author{T.~Inagaki}\affiliation{\SOKENDAI}\affiliation{\KEK} 
\author{S.~Ishibashi}\affiliation{\SAGA}
\author{N.~Ishihara}\affiliation{\KEK}
\author{H.~Ishii}\affiliation{\OSAKA} 
\author{E.~Iwai}\affiliation{\OSAKA}
\author{T.~Iwata}\affiliation{\YAMAGATA} 
\author{I.~Kato}\affiliation{\YAMAGATA} 
\author{S.~Kobayashi}\affiliation{\SAGA}
\author{S.~Komatsu}\affiliation{\OSAKA}
\author{T.~K.~Komatsubara}\affiliation{\KEK} 
\author{A.~S.~Kurilin}\affiliation{\DUBNA} 
\author{E.~Kuzmin}\affiliation{\DUBNA}
\author{A.~Lednev}\affiliation{\IHEP}\affiliation{\CHICAGO} 
\author{H.~S.~Lee}\affiliation{\PUSAN} 
\author{S.~Y.~Lee}\affiliation{\PUSAN} 
\author{G.~Y.~Lim}\affiliation{\KEK}
\author{J.~Ma}\affiliation{\CHICAGO}
\author{T.~Matsumura}\affiliation{\NDA}
\author{A.~Moisseenko}\affiliation{\DUBNA}
\author{H.~Morii}\altaffiliation{Present address: \KEK}\affiliation{\KYOTO}
\author{T.~Morimoto}\affiliation{\KEK}
\author{Y.~Nakajima}\affiliation{\KYOTO}
\author{T.~Nakano}\affiliation{\RCNP} 
\author{H.~Nanjo}\affiliation{\KYOTO}
\author{N.~Nishi}\affiliation{\OSAKA}
\author{J.~Nix}\affiliation{\CHICAGO}
\author{T.~Nomura}\altaffiliation{Present address: \KEK}\affiliation{\KYOTO}
\author{M.~Nomachi}\affiliation{\OSAKA}
\author{R.~Ogata}\affiliation{\SAGA}
\author{H.~Okuno}\affiliation{\KEK}
\author{K.~Omata}\affiliation{\KEK}
\author{G.~N.~Perdue}\altaffiliation{Present address: \NY}\affiliation{\CHICAGO} 
\author{S.~Perov}\affiliation{\DUBNA}
\author{S.~Podolsky}\altaffiliation{Present address: \GOMEL}\affiliation{\DUBNA}
\author{S.~Porokhovoy}\affiliation{\DUBNA}
\author{K.~Sakashita}\altaffiliation{Present address: \KEK}\affiliation{\OSAKA} 
\author{T.~Sasaki}\affiliation{\YAMAGATA} 
\author{N.~Sasao}\affiliation{\KYOTO}
\author{H.~Sato}\affiliation{\YAMAGATA}
\author{T.~Sato}\affiliation{\KEK}
\author{M.~Sekimoto}\affiliation{\KEK}
\author{T.~Shimogawa}\affiliation{\SAGA}
\author{T.~Shinkawa}\affiliation{\NDA}
\author{Y.~Stepanenko}\affiliation{\DUBNA}
\author{Y.~Sugaya}\affiliation{\OSAKA}
\author{A.~Sugiyama}\affiliation{\SAGA}
\author{T.~Sumida}\altaffiliation{Present address: \CERN}\affiliation{\KYOTO}
\author{S.~Suzuki}\affiliation{\SAGA}
\author{Y.~Tajima}\affiliation{\YAMAGATA}
\author{S.~Takita}\affiliation{\YAMAGATA} 
\author{Z.~Tsamalaidze}\affiliation{\DUBNA}
\author{T.~Tsukamoto}\altaffiliation{Deceased.}\affiliation{\SAGA} 
\author{Y.~Wah}\affiliation{\CHICAGO}
\author{H.~Watanabe}\altaffiliation{Present address: \KEK}\affiliation{\CHICAGO}
\author{M.~L.~Wu}\affiliation{\TAIWAN}
\author{M.~Yamaga}\altaffiliation{Present address: \RIKEN}\affiliation{\KEK}
\author{T.~Yamanaka}\affiliation{\OSAKA}
\author{H.~Y.~Yoshida}\affiliation{\YAMAGATA}
\author{Y.~Yoshimura}\affiliation{\KEK}
\author{Y.~Zheng}\affiliation{\CHICAGO}

\collaboration{E391a Collaboration}\noaffiliation

\begin{abstract}
We performed a search for the decay $\kggg$ with the E391a detector at KEK.
In the data accumulated in 2005, no event was observed in the signal region.
%with an estimated background of $0.19\pm0.93$ events.
Based on the assumption of $\kggg$ proceeding via parity-violation,
we obtained the single event sensitivity to be $\ses$,
and set an upper limit on the branching ratio to be $7.4\times10^{-8}$ at the $90\%$ confidence level.
This is a factor of $3.2$ improvement compared to the previous results.
The results of $\kggg$ proceeding via parity-conservation were also presented in this paper.
\end{abstract}
%
%================================================
%   Preprint numbers
%================================================
\maketitle
%================================================
%   Introduction
%================================================
We report the first results of a search for the decay $\kggg$
since the last experimental update in 1995~\cite{barr}.  
Although the decay is forbidden by charge-conjugation invariance,
it can proceed via weak parity-violating interactions without violating CP.
But due to further suppressions by the gauge invariance and Bose statistics~\cite{yang},
the branching ratio (BR) of $\kggg$ is expected to be very small.
Assuming the decay proceeds via $K_L^0 \rightarrow \pi^0\pi^0\gamma \rightarrow 3\gamma$ process
with two $\pi^0$'s internally converting to photons,
the calculated BR($\kggg$) is $3\times10^{-19}$~\cite{heiliger}.
Recently, 
a new calculation based on parity-violating model 
showed that the BR should be in the range of 
$7\times10^{-17}$ $\le$ BR($\kggg$) $\le$ $1\times10^{-14}$~\cite{jusak}.

%================================================
%  KEK E391a
%================================================
%================================================
%    Beamline
%================================================
%The $\kggg$ decay is studied at the KEK E391a experiment~\cite{e391run3}
%from neutral kaons which were produced by 12~GeV protons
%incident on a 0.8-cm-diameter and 6-cm-long platinum target.
The E391a experiment~\cite{website,e391run3} was conducted at KEK 
using neutral kaons produced by 12~GeV protons 
incident on a 0.8-cm-diameter and 6-cm-long platinum target.
The proton intensity was typically $2 \times 10^{12}$
per spill coming every 4 sec.
The neutral beam~\cite{beamline}, 
with a solid angle of 12.6~$\mu$str,
was defined by a series of six sets of collimators and a pair of sweeping magnets 
aligned at a production angle of 4~degrees. 
A 7-cm-thick lead block and a 30-cm-thick beryllium block were placed
between the first and second collimators
to reduce beam photons and neutrons.
%The beam size at $11.8$~m downstream of the target,
The beam size at the entrance of the detector (11.8~m downstream of the target), 
which was measured with the E391a detector by reconstructing the $\kppp$ decay,
was $3.7$ cm (FWHM) including the effects of detector resolution.
The beam line was kept in vacuum 
at 1 Pa after 5~m downstream of the target
and $1 \times 10^{-5}$ Pa inside the fiducial decay region.
The $K^0_L$ momentum measured at the entrance of the detector peaked around 2~GeV/$c$.
%11.8~m downstream of the target.

\begin{figure}[!bt]
   \includegraphics[angle=-90, width=8.6cm]{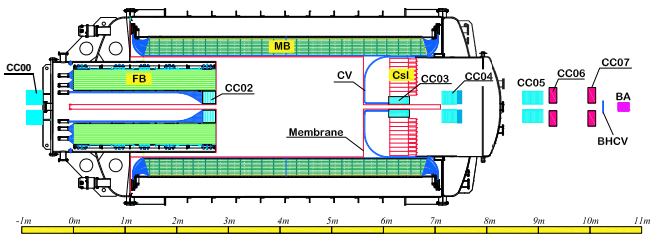}
   \caption{\label{fig:det1}%
	%(color online) 
	Schematic cross-sectional view of the E391a detector.
	``0m'' in the scale corresponds to 
	the entrance of the front barrel (FB) detector.
	$K^0_L$'s entered from the left side.
	}
\end{figure}

%================================================
%    Detectors
%================================================
Figure~\ref{fig:det1} shows a cross-sectional view of the E391a detector
and defines the origin of the coordinating system.
%\begin{figure*}[hbt]
%\end{figure*}
%
 The detector components were cylindrically assembled along the beam axis. 
 The electromagnetic calorimeter, labeled ``CsI'',
 measured the energy and position of the photons from $K_L^0$ and $\pi^0$ decays.
 It consisted of 496 blocks of \(7 \times 7 \times 30~\mbox{cm}^3\)
 undoped CsI crystal and 80 specially shaped CsI blocks used in the peripheral region, 
 covering a circular area with a 95 cm radius.
 In order to allow beam particles to pass through, 
 %the calorimeter had a $12 \times 12$~cm$^2$ beam hole at the center.
 there was a $12 \times 12$~cm$^2$ beam hole located at the center of the calorimeter.
 The main barrel (MB)~\cite{mb} and front barrel (FB) counters consisted of 
 alternating layers of lead and scintillator sheets with 
 total thicknesses of 13.5~$X_0$ and 17.5~$X_0$, respectively.
 To identify charged particles entering the calorimeter, 
 %a plastic scintillation counter (CV)
 an array of plastic scintillation counters (CV) 
 with a $12\times12$~cm$^2$ beam hole at the center
 was placed 50~cm upstream of the calorimeter.
 %hermetically covering the front of the calorimeter.
 Multiple collar-shaped photon counters
 (CC00, CC02--07) were placed along the beam axis to detect particles escaping in the beam direction.
 The CC02 was located at the upstream end of the $K^0_L$ decay region.
 The CC03 filled the volume between the beam hole 
 and the innermost layers of the CsI blocks in the calorimeter. 
 The vacuum region was separated by a thin multi-layer film (``membrane'') 
 between the beam and detector regions.
 Detailed descriptions of the E391a detector are given in~\cite{e391run3,detector}.
%================================================
%   Data period and statistics
%================================================

In this analysis, 
we used the data taken in the periods
Run-II (Feb.--Apr. 2005) and Run-III (Oct.--Dec. 2005) of E391a.
%Data were taken with a hardware trigger requiring two or 
%more shower clusters in the calorimeter with a cluster energy $\ge 60$~MeV. 
A hardware-based trigger system was used for data-taking, 
which required two or more shower clusters in the CsI calorimeter
with a cluster energy larger than $60$ MeV.
%We also required no activity in the CV and in some other photon counters. 
We imposed online cuts on the CV and some other photon counters. 
%================================================
%   Data Analysis - Event reconstruction
%================================================
The $K_L^0$ decays were simulated using the GEANT3 Monte Carlo (MC) generator~\cite{geant}
and were overlaid with accidental hits taken 
with a target-monitor trigger.
Since the decay of $\kggg$ is via weak interactions, 
parity conservation is not guaranteed.
There were three different models considered for the simulations:
the phase space, the parity-violating~\cite{pv}, and the parity-conserving~\cite{pc} interactions.

Candidates of $\kggg$ were selected by requiring three photon-like clusters in the CsI calorimeter 
without any in-time hits in the other detectors.
All clusters were required to be between 25 cm to 88 cm from the center of the beamline.
An additional selection criterion on the transverse momentum of $\kggg$ candidates,
$P_T$~$<$~0.05~GeV/$c$,
was required to suppress the $\kpp$ and $\kppp$ background events with undetected photons.
The decay vertex of $K_L^0$ candidates was calculated 
by requiring three photons to form the $K_L^0$ mass
and by constraining the vertex to lie along the beam axis.
The MC showed that 
$15\%$ of the well-reconstructed $\kggg$ events decayed before the exit of CC02 ($z=275$ cm).
To preserve acceptance, 
the fiducial decay Z-vertex ($Z_{K_{L}^{0}}$) region was defined to be between 200 and 550 cm.

%================================================
% BG suppression
%================================================
Most of the backgrounds to the decay $\kggg$ were related to 
%the inefficiency of veto detectors 
the detection inefficiency of photon counters or fusion clusters in the CsI calorimeter.
A fusion cluster is defined by two or more photons which are reconstructed together as a single cluster.
In previous E391a analyses,
a tight energy threshold was applied to the MB detector to reduce the detection inefficiency.
This caused a major signal acceptance loss due to splash-back and electromagnetic shower leakage
from the CsI calorimeter to the MB.
According to the MC simulations,
%true 
the undetected photons in the MB mostly entered the upstream region,
while splash-back and electromagnetic shower leakage tended to enter the downstream region.
%Therefore,
Thus,
a tighter energy threshold was applied to the upstream region of the MB to improve the detection efficiency,
and a looser threshold was applied to the downstream region to keep the signal acceptance.
%in this analysis
%the acceptance lose of the signal.
The particle-hit position on the MB was reconstructed 
using the TDC information measured from both ends of the MB counter.
The calibration and the simulation of TDC timing were carefully treated counter by counter.
%and run-dependently.
%
\begin{figure}[!h]
\begin{overpic}[width=8.6cm]{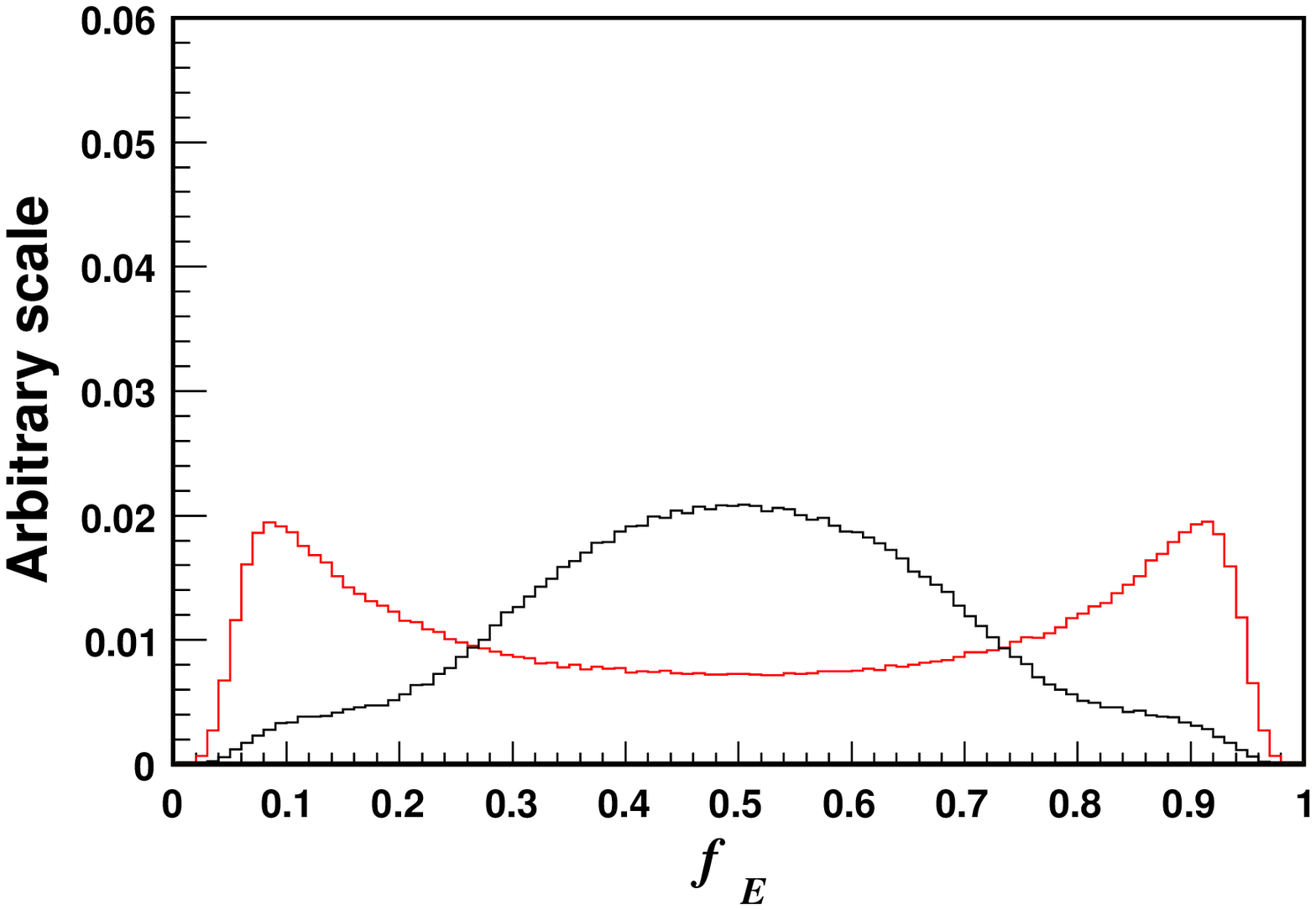}
\put(102,68){\includegraphics[width=4.5cm]{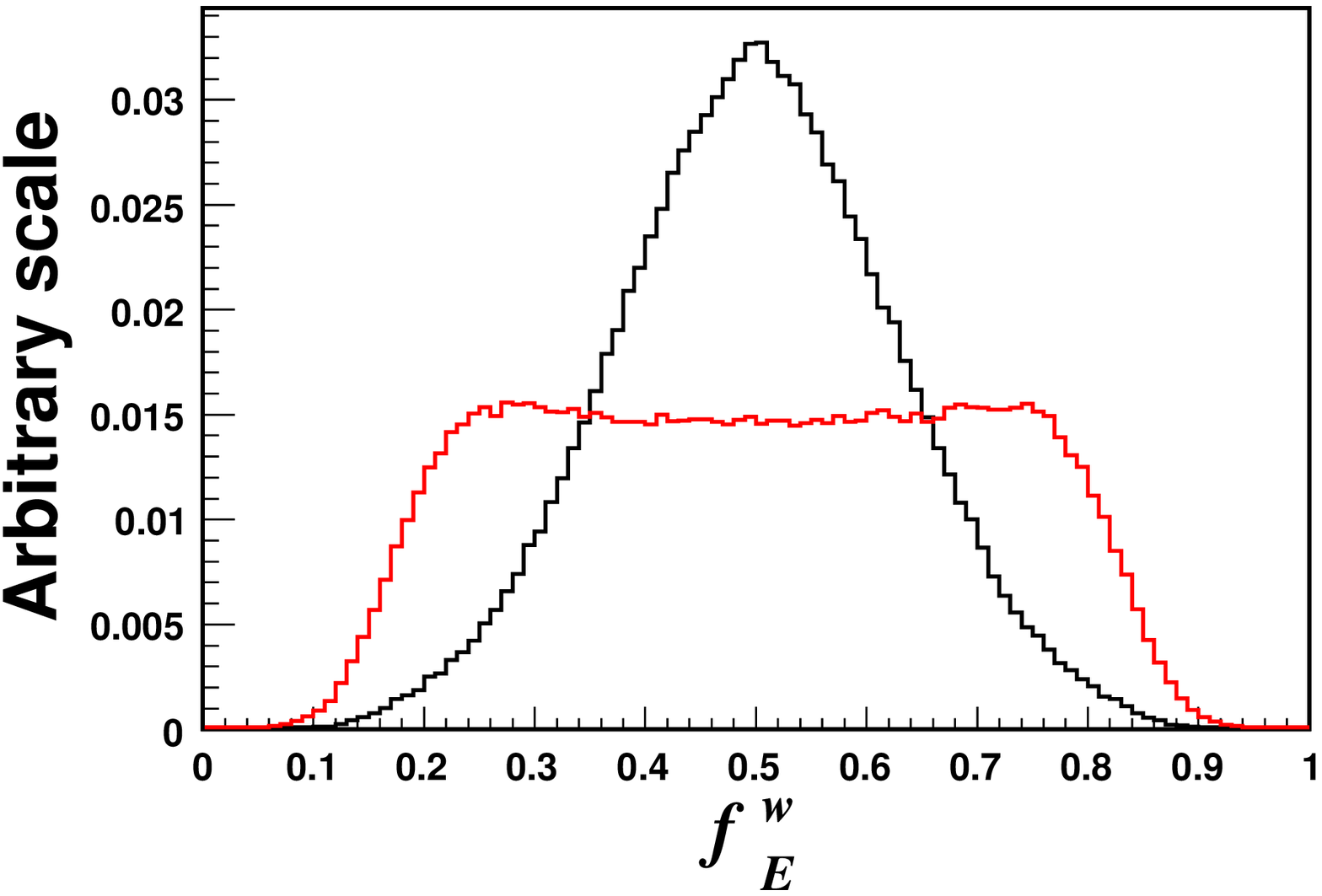}}
\end{overpic}
\caption{\label{fig:bcut}
Distributions of $f_E$ and $f_E^w$ (sub-figure) in the region of $x'>0$.
The values of $f_E$ and $f_E^w$ were normalized 
by the total sum of $f_E$ and $f_E^w$ in all regions, respectively.
The red histogram shows the distribution of single-photon clusters 
and the black histogram shows the distribution of fusion clusters.
A significant difference between the distributions in the region of $y'>0$ was also observed.}
\end{figure}
The fusion clusters were mainly suppressed using the neural network (NN) multivariate method trained 
%by single-photon and fusion clusters selected from $\kppp$ MC samples.
by the single-photon and the fusion clusters selected from $\kppp$ MC samples.
For further suppression,
the cluster was divided into four regions 
by a radial ($x'$) and a transverse ($y'$) line crossed at the center of energy of the cluster.
The MC study showed that the energy fractions, 
$f_E=\Sigma_i E_i~(i=crystals~in~the~cluster)$,
and the weighted energy fraction, $f_E^w=\Sigma_i E_i \times q'^2~(q'=x'~or~y')$, 
in these defined regions
were significantly different between the single-photon and the fusion clusters 
as shown in Fig.~\ref{fig:bcut}.
By applying cuts on $f_E$, $f_E^w$, and NN fusion together,
the dominant background source, 
fusion-related $\kppp$ events, were completely rejected in the MC.

\begin{figure}[!h]
   \includegraphics[width=8.6cm]{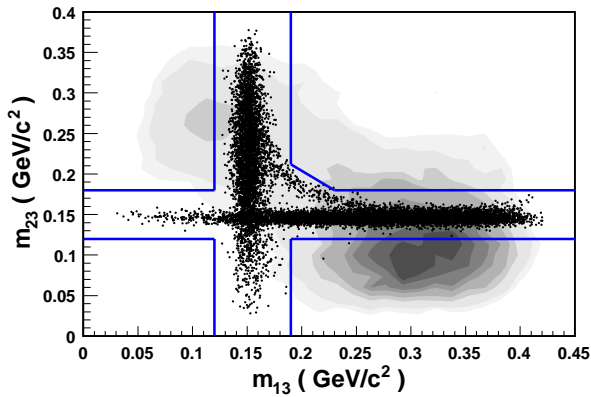}
   \caption{\label{fig:k2pi0}%
   %(color online)
   Distribution of $m_{13}$--$m_{23}$ of the MC events with three clusters in the CsI calorimeter.
   The contour shows the $\kggg$ parity-violating results, and the dots show the $\kpp$ results.
	 The events inside the cross region were rejected. 
	 %Four regions in the corners were defined as the signal region.	 
}
\end{figure}

Although the reconstructed $Z_{K_L^0}$ of the $\kpp$ events
with undetected photons was not precisely measurable,
MC studies on all types of $\kpp$ background events showed that
%the $Z_{K_L^0}^{measured}-Z_{K_L^0}^{true}$ distribution
the difference of the measured $Z_{K_L^0}$ and the true $Z_{K_L^0}$
had a mean of only 20~cm and a deviation of 10~cm.
%Since the difference between the measured and the true $Z_{K_L^0}$ is small,
Since this difference is small,
the measured $Z_{K_L^0}$ was used 
to reconstruct the invariant mass of the $i^{th}$ and $j^{th}$ photons, $m_{ij}$,
where the photons were sorted by carried energies. 
Events were then rejected if the reconstructed $m_{ij}$ matched the mass of pion, $m_{\pi^0}$.
There were three possible combinations to form the $m_{ij}$, 
and the relatively significant $m_{\pi^0}$ peaks were observed 
in the $m_{23}$ and $m_{13}$ distributions.
After rejecting the events with the values of $m_{12}$ matching $m_{\pi^0}$,
the $m_{13}-m_{23}$ distribution of the MC is shown in Fig.~\ref{fig:k2pi0}.
%The events inside the cross region were rejected.
%Four regions in the corners were defined as the signal region.
%Events inside the cross region were further rejected after the $m_{12}$ requirement, 
Events inside the cross region were further rejected, 
and then the four corner regions were defined as the signal region.

\begin{figure}[!h]
\begin{overpic}[width=8.6cm]{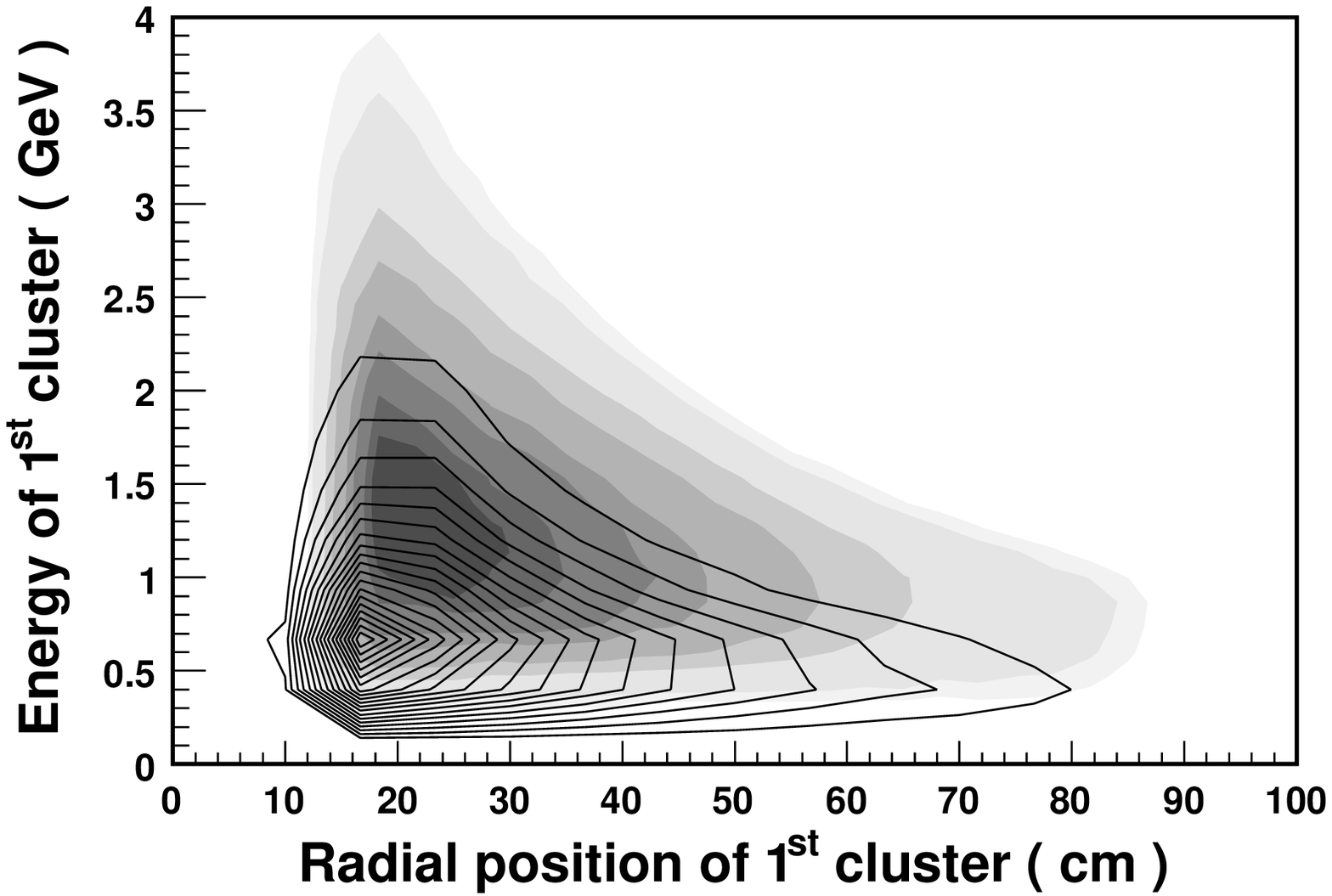}
%\put(3,28){\huge \LaTeX}
%\put(34,5){\includegraphics[scale=.07]{k3pi0.eps}}
\put(102,68){\includegraphics[width=4.5cm]{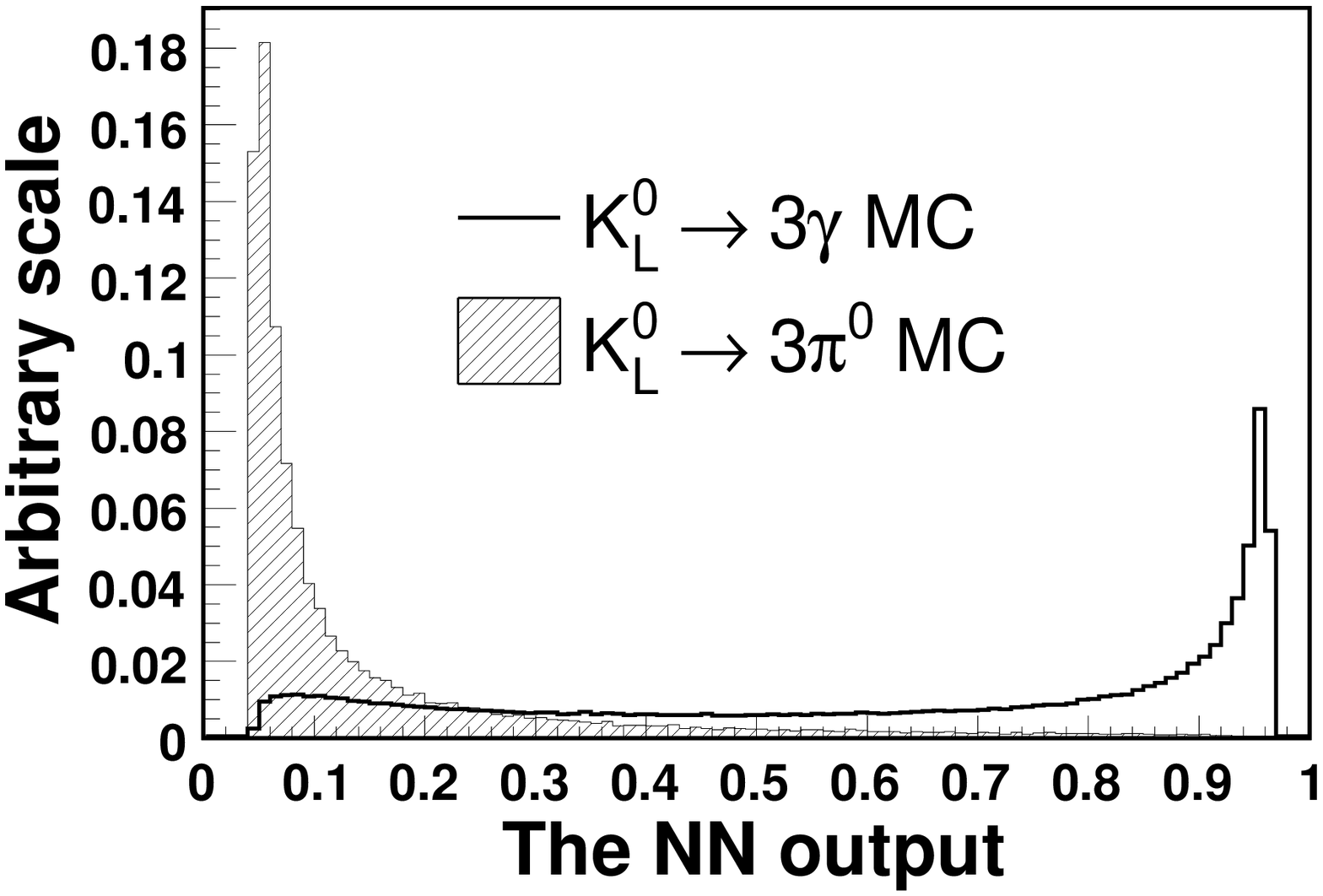}}
\end{overpic}
\caption{\label{fig:k3pi0}%
   %(color online)
   Distribution of Cluster radial position vs. cluster energy of the MC events
   and the distributions of the NN output (sub-figure). 
   The banded contour shows the $\kggg$ parity-violating results, 
   and the line contour shows the $\kppp$ results.
	 The significant differences in the distributions were also observed in the other two clusters.
}
\end{figure}

%The $\kppp$ mode was the dominant background source.
In the MC study,
the $\kppp$ and $\kggg$ events showed different behaviors
in the distributions of cluster energy and position.
This was due to the fact that the number of photons appearing 
in the final state of the two processes were different.
%These variables, which relied on the CsI measurements in addition to $Z_{K_L^0}$, 
These variables, which relied on the measurements from the CsI calorimeter, 
and $Z_{K_L^0}$ were combined together by using the NN method,
and the results are shown in Fig.~\ref{fig:k3pi0}.
After requiring the events with the NN output larger than $0.7$,
all the remaining 68 $\kppp$ MC events were rejected and $62.9\%$ of the $\kggg$ MC events remained.
%These results were estimated basing on a loose cut set due to the lack of the $\kppp$ MC statistics.
%
%\begin{overpic}[scale=.25,unit=1mm]{k3pi0.eps}
The $\kgg$ decay could contribute to the background 
if an accidental cluster arrived in the calorimeter in-time with the event.
Since the energies of accidental clusters were relatively small compared to decayed photons, 
the $\kgg$ was easily identified by requiring two higher energy clusters.  
The center of energy of the two decayed photons should distribute around the beam center, 
with combined $P_T$ equal to that of their parent $K_L^0$. 
Thus, the event was rejected if the center of energy of the two highest energy clusters 
was less than $4$ cm from the beam center.

\begin{figure}[!h]
   \includegraphics[width=8.6cm]{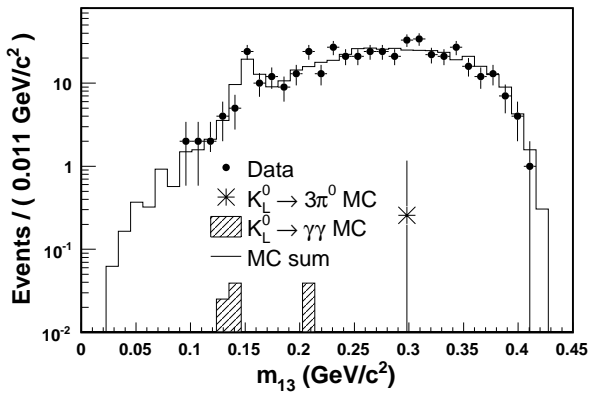}
   \caption{\label{fig:m13}%
   %(color online)
   Distributions of $m_{13}$ of the events in the normalization region
   (0.13 GeV/$c^2$ $<$ $m_{12}$ $<$ 0.17 GeV/$c^2$).
   The points with error bars show the data, 
   the star sign shows the only survived $\kppp$ MC event, 
   and the shaded histograms show the $\kgg$ MC events.
   The hollow solid histogram is the sum of the $\kppp$, $\kpp$ and $\kgg$ MC results.  
}
\end{figure}

With all cuts applied,
$3$ events of $4\times10^{9}$ generated $\kpp$ MC events remained in the signal region as shown in Fig.~\ref{fig:k2pi0},
corresponding to $0.16\pm0.10$ events after normalization.
We generated $3.2\times10^{10}$ $\kppp$ MC events ($104\%$ of data) for Run-II 
and $7\times10^{10}$ MC events ($322\%$ of data) for Run-III.
No events passed the $\kggg$ cuts.
For the $\kgg$ MC,
only one in $4\times10^{9}$ generated events passed the cuts,
corresponding to $0.03^{+0.05}_{-0.03}$ events after normalization.
For the MC with no event passing the cuts,
the number of the survived events was conservatively set to be $0^{+1}_{-0}$.
The total number of expected background events from the three sources was 
then estimated to be $0.19^{+0.93}_{-0.10}$,
where the quoted error includes statistical and systematic uncertainties.
%The statistical error of the MC with no events passing the cuts was conservatively set to be $1$.
The total error is dominated by the $\kppp$ statistical uncertainty.
The MC events were normalized using the number of data events 
in the region of 0.13~GeV/$c^2$~$<$ $m_{12}$~$<$~0.17~GeV/$c^2$ (normalization region).
\begin{table}[t]
   \caption{\label{table:BG}%
	Summary of the estimated numbers of the backgrounds.
	The quoted errors include statistical and systematic uncertainties.
	}
   \begin{ruledtabular}
   \begin{tabular}{lccc}
Mode		&	Run-II	& Run-III	& Total \\
\hline
%$\kpp$	&	$0.12\pm0.09$	&	$0.04\pm0.04$	&	$0.16\pm0.10$\\
%$\kppp$	&	$0.00\pm0.88$	&	$0.00\pm0.26$	&	$0.00\pm0.92$\\
%$\kgg$	&	$0.00\pm0.04$	&	$0.03\pm0.03$	&	$0.03\pm0.05$\\
$\kpp$	&	$0.12\pm0.09$&	$0.04\pm0.04$	&	$0.16\pm0.10$\\
$\kppp$	&	$0.00^{+0.88}_{-0.00}$		&	$0.00^{+0.26}_{-0.00}$	&	$0.00^{+0.92}_{-0.00}$\\
$\kgg$	&	$0.00^{+0.04}_{-0.00}$	&	$0.03\pm0.03$	&	$0.03^{+0.05}_{-0.03}$\\
   \end{tabular}
   \end{ruledtabular}
\end{table}
\begin{table*}[htpb]%[hb]
\caption{\label{table:Results}%
Summary of the acceptances of the $\kggg$ decay ($A$($\kggg$)), 
combined single event sensitivities ($SES_{combined}$),
and the upper limits (UL) at the $90\%$ confidence level for different $\kggg$ decay models.
}
\begin{center}
\begin{ruledtabular}
\begin{tabular}{lccc}
Decay Model	&	$A$($\kggg$)	& $SES_{combined}$	&	UL \\
\hline
Phase space					&	$(0.99\pm0.01)\times10^{-4}$	&	$(3.75\pm0.16)\times10^{-8}$ 	& $8.62\times10^{-8}$ \\
Parity violation		&	$(1.15\pm0.02)\times10^{-4}$	&	$(3.23\pm0.14)\times10^{-8}$	& $7.42\times10^{-8}$ \\
Parity conservation	&	$(1.11\pm0.02)\times10^{-4}$	&	$(3.28\pm0.14)\times10^{-8}$	& $7.54\times10^{-8}$ \\
\end{tabular}
\end{ruledtabular}
\end{center}
\end{table*} 
The $m_{13}$ distributions of the data and the MC results in this region are shown in Fig.~\ref{fig:m13}.
The region of 
$0.13$ GeV/$c^2$ $<$ $m_{13}$ $<$ $0.18$ GeV/$c^2$ and $m_{12}$ $>$ $0.18$ GeV/$c^2$
was defined as the upper sideband. 
The region of 
$0.13$ GeV/$c^2$ $<$ $m_{13}$ $<$ $0.18$ GeV/$c^2$ and $m_{12}$ $<$ $0.12$ GeV/$c^2$
was defined as the lower sideband.
In the full data,
we observed $164$ events with a MC prediction of $158.9 \pm 8.2$ events in the upper sideband region
and $6$ events with a prediction of $7.1 \pm 1.2$ events in the lower sideband region.
Since the results of the three background sources
well described the data in both normalization region and sidebands,
other background sources, such as neutron interactions, were neglected.
The estimated backgrounds are summarized in Table~\ref{table:BG}.

\begin{figure}[!h]
   \includegraphics[width=8.6cm]{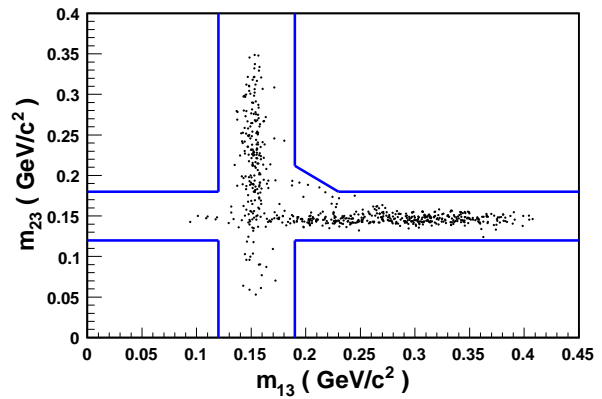}
   \caption{\label{fig:final}%
   %(color online)
   Distribution of $m_{13}$--$m_{23}$ of the data with all selection cuts imposed.
   %Four boxes in the corners indicate the signal region for $\kggg$. 
   No event was observed in the signal region.
}
\end{figure}

With all selection cuts applied to the data, no events survived in the signal region (Fig.~\ref{fig:final}).
The single event sensitivity for $\kggg$ was defined as
\begin{eqnarray}
	\nonumber
    SES(\kggg) = \frac{1}{A(\kggg)\cdot N(K^0_L~\mbox{decays})},
\end{eqnarray}
%\iffalse
where $A$($\kggg$) is the acceptance for $\kggg$ and $N$($K^0_L$ decays) is the integrated $K_L^0$ flux.
The $K_L^0$ flux was evaluated by the $\kpp$ mode and was cross-checked by the $\kppp$ mode.
The $K_L^0$ fluxes at 10 m from the target were determined to be
$(1.57\pm0.09)\times10^{11}$ for Run-II and $(1.11\pm0.07)\times10^{11}$ for Run-III
based on the number of decays downstream of that point.
The quoted error in the $K_{L}^{0}$ flux combines statistical and systematic uncertainties.
Systematic uncertainties from disagreements between data and the MC simulation dominated the total error.
% 
%The SES's for Run-II and Run-III were combined using
%\begin{eqnarray}
%	\nonumber
%    \frac{1}{SES_{combined}} = \frac{1}{SES_{\tiny{\mbox{Run-II}}}}+\frac{1}{SES_{\tiny{\mbox{Run-III}}}}.
%\end{eqnarray}
%
The $\kggg$ acceptance varied with decay models.
Results from three models are summarized in Table~\ref{table:Results}: 
phase space, parity-violating and parity-conserving interactions.
The upper limits at the $90\%$ confidence level were calculated based on Poisson statistics.
The parity-violating model (CP conserved) was used to obtain the final result 
and set an upper limit of
the BR$\left(\kggg\right)$ to be $7.4\times10^{-8}$ at the $90\%$ confidence level.
This is a factor of 3.2 improvement over the previous results

% -----------------------------------------------------------------------
%  Acknowledgements
% -----------------------------------------------------------------------
\begin{acknowledgements}
We are grateful to the operating crew of the KEK 12-GeV proton synchrotron 
for their successful beam operation during the experiment. 
This work has been partly supported by a Grant-in-Aid from the MEXT and JSPS in Japan, 
%a grant from National Science Council in Taiwan, 
a grant from National Science Council and Ministry of Education in Taiwan, 
from the U.S. Department of Energy and from Korea Research Foundation.
\end{acknowledgements}

\bibliographystyle{plain}

\end{document}